\documentclass[11pt]{article}
\usepackage{amssymb}
\usepackage{amsmath}
\usepackage{graphicx}
\usepackage{hep}
\usepackage{sbl_th}
\newcommand\pubblock{\rightline{\begin{tabular}{l} \pubnumber\\
      \end{tabular}}}
\usepackage{amssymb, axodraw}
\usepackage{amsmath}
\usepackage{graphicx}
\usepackage{hep}
\newcommand{\vertex}{\Pphoton\Pphoton h}
\newcommand{\twophoton}{\Pphoton\Pphoton}
\newcommand{\hzero}{\ensuremath{\PHiggslightzero}} 
\newcommand{\Hzero}{\ensuremath{\PHiggsheavyzero}} 
\newcommand{\HSM}{\ensuremath{\PHiggsheavy}} 
\newcommand{\Azero}{\ensuremath{\PHiggspszero}} 
\newcommand{\Hpm}{\ensuremath{\PHiggspm}} 

\newcommand{\bsg}{\ensuremath{\mathcal{B}(b \to s \gamma)}}
\newcommand{\pdirect}{\HepProcess{\Pphoton\Pphoton \to h}}
\newcommand{\pdirecth}{\HepProcess{\Pphoton\Pphoton \to \hzero}}

\newcommand{\pdirectA}{\HepProcess{\Pphoton\Pphoton \to \Azero}}

\newcommand{\pfusion}{\HepProcess{\APelectron\Pelectron \to \APelectron\Pelectron\Pphoton^*\Pphoton^* \to \APelectron\Pelectron + h}}
\newcommand{\pminifusionh}{\HepProcess{\APelectron\Pelectron \to \APelectron\Pelectron\hzero}}

\newcommand{\pp}{$\gamma\gamma\,\,$}
\newcommand{\epem}{\APelectron\Pelectron}
\newcommand\pubnumber{UB-ECM-PF-09/06}
\newcommand{\pmzero}{$\PHiggs^+\PHiggs^-\PHiggslightzero\,\,$}
\newcommand{\newtext}[1]{\text{#1}}

\begin{document}
\pubblock

\begin{center}
\vspace{1.3cm}
\begin{center}
\vskip 1.5cm
    {\large \textsc{Single Higgs-boson production through $\gamma\gamma$
    scattering\\
within the general 2HDM}}

    \vskip 0.7cm

\textbf{Nicol\'as Bernal, David L\'opez-Val, Joan Sol\`a}

\vskip 0.7cm

\textit{High Energy Physics Group, Dept. ECM, and Institut de Ci{\`e}ncies del Cosmos\\
Univ. de Barcelona, Av. Diagonal 647, E-08028 Barcelona, Catalonia, Spain}\\

\vskip5mm

E-mails: bernal@ecm.ub.es, dlopez@ecm.ub.es, sola@ecm.ub.es.
\vskip15mm

\end{center}

\begin{quotation}
\noindent {\large\it \underline{Abstract}}.\ \ The production of a
single neutral Higgs boson $h$ through (loop-induced) $\gamma\gamma$
collisions is explored in the context of the linear colliders within
the general Two-Higgs-Doublet Model (2HDM). Two different mechanisms
are analyzed: on the one hand, the scattering $\gamma\gamma\to h$ of
two real photons in a $\gamma\gamma$ collider; on the other, the
more traditional mechanism of virtual photon fusion, $\pfusion$.
Owing to the peculiar properties of the Higgs boson
self-interactions within the general 2HDM,  we find that the overall
production rates can be boosted up significantly, provided the
charged Higgs mass is not too heavy. For example, if
$M_{\Hpm}\gtrsim 100\,GeV$ and, in addition, $M_{\PHiggslight^0}$
falls in the ballpark of the LEP bound on the SM Higgs mass up to a
few hundred GeV, the cross-sections may typically render
$\langle\sigma_{\gamma\gamma\to h}\rangle\sim 0.1-1\,\picobarn$ and
$\sigma(\pminifusionh)\lesssim 0.01\,\picobarn$ -- in both cases
well above the SM prediction. Although for $M_{\Hpm}> 300\,GeV$ the
rates become virtually insensitive to the Higgs boson
self-couplings, a significant tail of non-SM effects produced by the
combined contribution of the Yukawa couplings and gauge bosons could
still reveal a smoking gun.
\end{quotation}
\vskip 8mm
\end{center}

\newpage

\section{Introduction}
\label{sec:intro}

The Higgs mechanism is the most fundamental lingering issue that
remains experimentally unsettled in Particle Physics. It is
difficult to overemphasize that this issue stands right in the core
of our present understanding of the Standard Model (SM). However, we
cannot exclude that the Higgs sector is larger than expected, the
most paradigmatic extension being the Minimal Supersymmetric
Standard Model (MSSM)\,\cite{MSSM}, which involves two doublets of
complex scalar fields. The physical spectrum consists of two charged
states, $H^{\pm}$, two neutral CP-even states $h^0, H^0$ (with
masses $M_{h^0}<M_{H^0}$) and one CP-odd state $A^0$\,\cite{hunter}.
Let us recall that the self-interactions of the SUSY Higgs bosons
are rather inconspicuous, in the sense that they cannot be enhanced
as compared to the ordinary gauge interactions and, therefore, do
not present a very distinctive phenomenology. The bulk of the
enhancing capabilities of the MSSM Lagrangian resides, instead, in
the rich structure of Yukawa couplings between Higgs bosons and
quarks or between quarks, squarks and chargino-neutralinos. The
stupendous phenomenological opportunities associated to these
supersymmetric structures are well-known since long ago
(cf.\,\cite{SUSYeffects1,SUSYeffects2}) and have been continuously
updated in the literature (for recent reviews, see e.g.
\,\cite{HiggsRevs}).

On the other hand, we should be prepared for alternative forms of
Higgs boson physics of a more generic kind, whose potential
implications can be equally outstanding and nevertheless be
concentrated on very different sectors of the model. This could e.g.
be the case of the general (unconstrained) Two-Higgs-Doublet Model
(2HDM), where again two doublets of complex scalar fields are
introduced, leading to a similar physical spectrum
$h^0,H^0,A^0,H^{\pm}$, but without being subdued by the severe
restrictions enforced by the supersymmetric transformations. The
result is a Higgs potential with a collection of Higgs boson
self-interactions which, in contrast to the SUSY case, can be highly
enhanced in comparison to the gauge couplings. We refer the reader
to Ref. \cite{hunter,HiggsRevs} for further details.

Let us assume that the LHC unveils a neutral Higgs boson. An
essential part of the very process of identification will be to
disclose just whether such particle is actually the neutral SM Higgs
boson, a neutral member of a SUSY extension of the SM (typically the
MSSM), or rather a generic neutral Higgs boson of a non-SUSY
alternative setup, such as e.g. the general 2HDM. In this task, the
complementary help of the future linear $e^+e^{-}$ colliders
(linac)\,\cite{ILCPhysics}, such as the ILC and CLIC, can play a
momentous role to unravel the ultimate nature of the purported Higgs
boson scalar particle(s) presumably produced at the LHC. There are
many studies in the literature supporting this fact. For instance,
the trilinear (3H) couplings have been investigated
phenomenologically in TeV-class linear colliders in
\cite{pairmssm,Djouadi:1999gv,Fawzy02,Arhrib:2008jp} through the
double-Higgs strahlung process $\APelectron\Pelectron \to
\PHiggsheavy \PHiggsheavy \PZ$ or the $\PW\PW$ double-Higgs fusion
mechanism $\APelectron \Pelectron \to \PHiggsplus\PHiggsminus
\Pnue\APnue$. Unfortunately, the cross-section turns out to be
rather small  both in the SM and in the MSSM. Quite in contrast, it
has recently been shown that the general 2HDM can provide
cross-sections two to three orders of magnitude larger within the
same experimental setup\,\cite{Ferrera:2007sp,Hodgkinson:2009uj}.

Closely connected to the physics of the linear colliders will be the
physics of the $\gamma\gamma$ colliders\,\cite{gammacolliders}. As
is well-known, a collider of this sort can be optionally realized
from a linac by the process of backward Compton scattering between
laser photons and the linac leptons. Not surprisingly, a clean
machine as a $\gamma\gamma$ collider should enable us to probe the
most sensitive theoretical structures of gauge theories, and
certainly the Higgs sector is a most preeminent one.

Currently, a renewed thrust of theoretical activity has been
invested in double Higgs production in $\gamma\gamma$ collisions
within the general 2HDM\,\cite{2phot_2hdm} -- see also
\cite{GrifolsPascual80,old_twophoton} for earlier related work, and
\cite{2phot_mssm,maria} for single and double Higgs production in
the SM and the MSSM. In this Letter, we wish to further explore the
Higgs boson self-interactions in the general 2HDM by focusing on the
process of single neutral Higgs boson production $\gamma\gamma\to
h=\hzero, \Hzero, \Azero$ in the context of both \pp real scattering
and \pp virtual fusion in \epem colliders (see
Fig.\,\ref{fig:process}).

\section{Loop-induced $\vertex$ interactions within the 2HDM:
general features and computational setup}

Let us recall that the general 2HDM \cite{hunter} is obtained by
canonically extending the SM Higgs sector with a second $SU_L(2)$
doublet carrying weak hypercharge Y = +1, so that it contains 4
complex scalar fields. The free parameters in the most general,
CP-conserving, 2HDM potential can be expressed in terms of the
masses of the physical Higgs particles, $M_{\PHiggslightzero}$,
$M_{\PHiggsheavyzero}$, $M_{\PHiggspszero}$, $M_{\Hpm}$, the ratio
$\tan\beta=v_2/v_1$ of the two VEV's giving masses to the up- and
down-like quarks, the mixing angle $\alpha$ between the two CP-even
states, and, finally, the coupling $\lambda_5$ which cannot be
absorbed in any of the previous quantities\,\footnote{Throughout the
paper, we use the notation and conventions of
Ref.~\cite{Ferrera:2007sp}, to which we refer the reader for further
details. Here, in contrast to that reference, we leave $\lambda_5$
as a fully independent parameter.}. In turn, the possible 2HDM
coupling patterns in the Higgs-fermion sector are commonly sorted
out as follows: i) type-I models, in which only one Higgs doublet
couples to fermions, whereas the other doublet does not; and ii)
type-II models, wherein a doublet couples only to down-like fermions
and the other doublet only to up-like fermions. In either way one
may avoid the appearance of dangerous (tree-level) Flavor Changing
Neutral Current (FCNC) processes\,\cite{hunter}. The MSSM Higgs
sector is actually a type-II one, but of a very restricted sort
(enforced by SUSY invariance)\cite{MSSM}.

On top of that a number of important restrictions, emerging from
either the available experimental data and the theoretical
consistency of the model, must be taken into account in order to
obtain a more realistic output. Although we have already described
these constraints in \cite{{Ferrera:2007sp},Hodgkinson:2009uj} (see
also \cite{WahabElKaffas:2007xd}), we will introduce some
qualifications here. To start with, there are (additional) stringent
constraints coming from (one-loop induced) low-energy FCNC
processes, mainly from the charged Higgs boson contributions to
$\bsg$ \cite{Misiak:2006zs}, which require $M_{\Hpm} > 295\,\GeV$
(for $\tan\beta\geqslant 1$) in type-II models. Let us emphasize
that this bound does not apply to type-I models since for them the
charged Higgs couplings to fermions are proportional to $\cot\beta$
and hence the loop contributions are highly suppressed at large
$\tan\beta$. Furthermore, the approximate $SU(2)$ custodial symmetry
severely restricts the radiative corrections to the $\rho$ parameter
from the 2HDM degrees of freedom; experimentally $|\delta\,\rho|
\leq 10^{-3}$\,\cite{Amsler:2008zzb}. Moreover, there are of course
the bounds stemming from the unsuccessful Higgs boson searches at
LEP and the Tevatron \cite{Amsler:2008zzb}. Besides, a very
important set of conditions emerges from the unitarity constraints.
A substantial number of studies are devoted to this subject in the
literature \cite{unitarity_main,unitarity2}, although their
conclusions are not always fully coincident. Alternatively, one can
stick to a less restrictive (albeit well-motivated) framework based
on requiring that none of the triple and quartic Higgs boson
self-couplings (in the mass-eigenstate basis) should be larger than
the upper value of the corresponding couplings in the SM.
Ultimately, this condition is grounded on the Lee-Quigg-Thacker
unitarity bound\,\cite{LQT} on the SM Higgs boson mass. In this way,
we obtain a double set of conditions that significantly harness the
size of the 3H and 4H couplings in the 2HDM:
\begin{eqnarray}
|C_{hhh}|\le \left|
\lambda_{\PHiggsheavy\PHiggsheavy\PHiggsheavy}^{(SM)}(M_{\HSM} \simeq 1 \,\TeV) \right|
=\left. \frac{3\,e\,M_{\HSM}^2}{2\,\sin\theta_W\,M_W}\right|_{M_{\HSM}=1
\,\TeV}\,, \label{eq:unitary3h}
\end{eqnarray}
\begin{eqnarray}
|C_{hhhh}|\le \left|
\lambda_{\PHiggsheavy\PHiggsheavy\PHiggsheavy\PHiggsheavy}^{(SM)}(M_{\HSM}
\simeq 1 \,\TeV) \right| =\left.
\frac{3\,e^2\,M_{\HSM}^2}{4\,\sin^2\theta_W\,M^2_W}\right|_{M_{\HSM}=1
\,\TeV} \label{eq:unitary4h}
\end{eqnarray}
($-e$ being the electron charge and $\theta_W$ the weak mixing
angle). In the following, we will discuss our results by taking into
account the conditions \eqref{eq:unitary3h} and
\eqref{eq:unitary4h}, and we will briefly compare them with the
restrictions derived in Ref.~\cite{unitarity_main}. Furthermore, we
shall impose that the EW vacuum is stable, which is tantamount to
say that we demand the quartic interaction terms in the potential
not to give negative contributions producing an unbounded potential
from below\cite{vacstab}. This condition leads to
\begin{figure}[t]
\begin{center}
\begin{tabular}{cc}
 \hspace{1cm}
\includegraphics[scale=1.5]{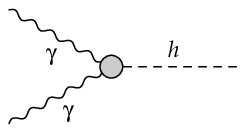}
\quad&
\includegraphics[scale=1.5]{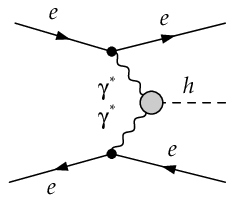}
 \\
\, & \, \\
(a) & (b)
\end{tabular}
\caption{\footnotesize{Generic Feynman diagrams describing the
single Higgs production process through the mechanisms of a) direct
scattering of a real photon pair and b) virtual $\twophoton$ fusion
in $\APelectron\Pelectron$ collisions. The grey blobs stand for the
generic loop-induced $\vertex$ interaction at any order in
perturbation theory.}} \label{fig:process}
\end{center}
\end{figure}
\begin{eqnarray}
&&\phantom{xxxxxxxxx}\Lambda_1>0,\ \ \ \ \ \
\Lambda_2>0,\nonumber\\
&&\sqrt{\Lambda_1\Lambda_2}+\Lambda_3+\text{Min}(0,\Lambda_4+\Lambda_5,
\Lambda_4-\Lambda_5)>0,
\end{eqnarray}
where the parameters $\Lambda_i$ are defined in terms of the
$\lambda_i$ ones\,\cite{Ferrera:2007sp} as follows:
\begin{eqnarray}
\Lambda_1&=&2\,(\lambda_1+\lambda_3),\ \ \ \
\Lambda_2=2\,(\lambda_2+\lambda_3), \ \ \ \
\Lambda_3=2\,\lambda_3+\lambda_4,\nonumber\\
\Lambda_4&=&-\lambda_4+\frac12\,(\lambda_5+\lambda_6),\ \ \ \ \
\Lambda_5=\frac12\,(\lambda_5-\lambda_6).
\end{eqnarray}

In this Letter, we are concerned with the production of a
Higgs-boson via \pp scattering. This process can proceed through the
following two basic and independent mechanisms:
\begin{itemize}
\item{Direct scattering of two real photons $\pdirect$, see
Fig.~\ref{fig:process}a};
\item{Virtual two-photon fusion in $\APelectron\Pelectron$ collisions,
namely $\pfusion$ (Fig.~\ref{fig:process}b}).
\end{itemize}

\begin{figure}[t]
\begin{center}
\includegraphics[scale=0.8]{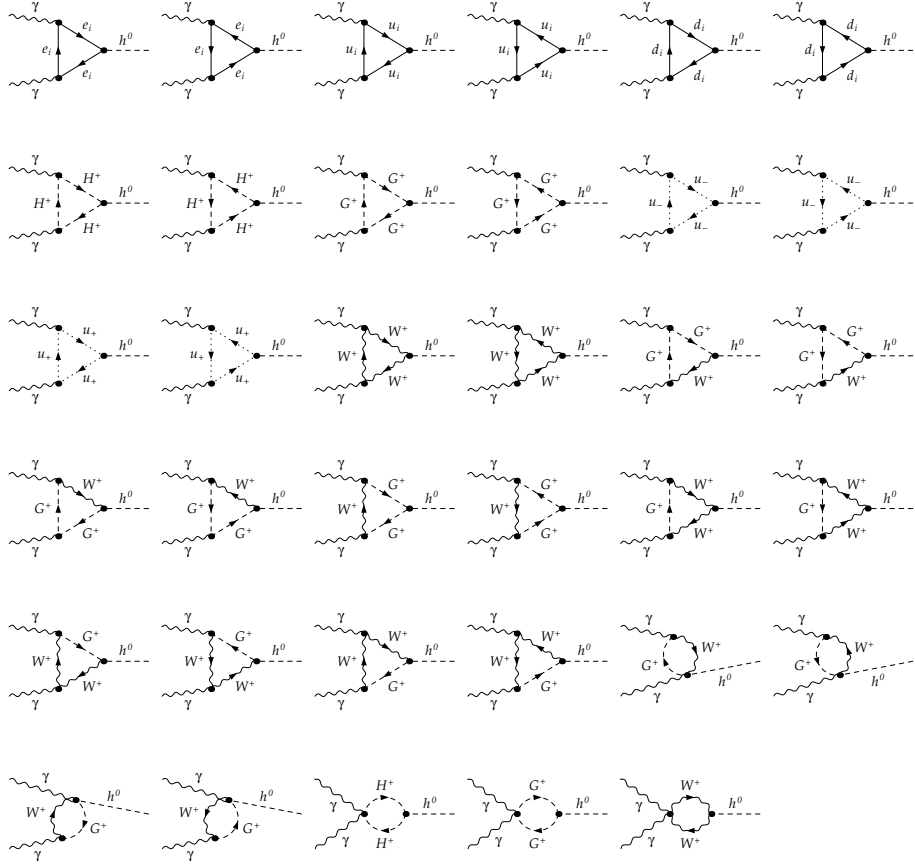}
\caption{\footnotesize{Feynman diagrams describing the process
$\pdirecth$, at the one-loop level, within the 2HDM. In the SM case,
we have to disregard the diagrams with charged Higgs bosons.}}
\label{fig:loop}
\end{center}
\end{figure}

\noindent Although a tree-level $\vertex$-coupling is not allowed by
the electromagnetic gauge symmetry, this interaction is generated at
the quantum level through a plethora of radiative corrections, whose
description in terms of Feynman diagrams is displayed in
Figure~\ref{fig:loop}. The entire set of diagrams corresponds to the
production process $\gamma\gamma\to h$ for both CP-even states
$h=h^0, H^0$, whereas for the production of the CP-odd one,
$\pdirectA$, only the first line of diagrams is allowed (owing to C
and CP-invariance).
The relevant cross-section can be written in terms of the partial
width of the decay $h\to\gamma\gamma$ as follows:
\begin{eqnarray}
{\sigma}(\pdirect) =\frac{8\,\pi^2}{M_h}\,\Gamma(\HepProcess{h
\to\Pphoton\Pphoton})\,\delta(s-M^2_h)\,(1+\eta_1\,\eta_2)=
8\pi\,\frac{\Gamma(h\to\gamma\gamma)\,\Gamma_h\,(1+\eta_1\eta_2)}{(s-M^2_h)^2
+ M_h^2\,\Gamma_h^2} \label{eq:generic2}\,,
\end{eqnarray}
with $\eta_{1,2}=\pm 1$ the helicities of the two colliding photons
and where we have used a standard representation for the
$\delta$-distribution. Following the above strategy, and computing
the diagrams contributing to the amplitude with the help of the
computational packages \emph{FeynArts}, \emph{FormCalc} and
\emph{LoopTools} \cite{feynarts}, we may finally arrive at the
$\gamma\gamma$-induced single Higgs boson cross-section. In order to
obtain more accurate results, a running value for the
electromagnetic coupling constant $\alpha_{em}(M_Z) = 1/127.9$ has
been used.

\section{Single Higgs boson production in a $\gamma\gamma$ collider}

The basic \pp collider operates mainly through the mechanism of
Compton (back)scattering of laser photons off the linac
beams\,\cite{gammacolliders}. In a nutshell: a photon from a laser
pulse collides with a high energy electron (or positron) at a small
angle; as a result, the electron recoils and one is left with a
Compton-scattered photon traveling in the direction of the original
incident electron. The efficiency of the $e^{\pm}\to\gamma$
``conversion'' will depend on many factors, such as the energy of
each of the beams, the properties of the laser as well as on several
non-linear effects and subleading mechanisms which turn out to
modulate the overall process. As a matter of fact, the entire
procedure comes down to furnish a $\gamma\gamma$ luminosity
spectrum, for which several standard parameterizations are
available. This spectrum is essential to compute the expected number
of events (in our case the number of produced Higgs bosons of a
particular neutral species $h=h^0,H^0,A^0$). To this end, the
cross-section computed in the previous section must be appropriately
folded with the normalized (dimensionless) photon densities provided
by the given parameterizations. The total (unpolarized)
$\gamma\gamma$ cross-section, after $e^{\pm}\to\gamma$
``conversion'' of the primary linac beam, can finally be engineered
from the following recipe:
\begin{eqnarray}
\langle\sigma_{\gamma\gamma\to h}\rangle(s) &=& \sum_{\{ij\}}
\int_{0}^1\,d\tau\,\frac{d\,\mathcal{L}_{ij}^{ee}}{d\tau}\,\hat{\sigma}_{\eta_i\,\eta_j}(\hat{s})\,,
\label{eq:sigmatotal}
\end{eqnarray}
where the partonic cross-section $\hat{\sigma}_{\eta_i\,\eta_j}$ is
given by
\begin{eqnarray}
\hat{\sigma}_{\eta_i\,\eta_j}(\hat{s}) =
\frac{\Gamma_h}{M_h}\,\frac{\big|\mathcal{M}_{\eta_i\,\eta_j}(\pdirect)\big|^2}{(\hat{s}-M_h^2)^2+M_h^2\Gamma_h^2}\,,
\label{eq:generic3}
\end{eqnarray}
\newtext{with $\hat{s}=\tau\,s$ ($s$ being the
CM energy of the primary linac machine)}. In the above expression,
$d\,\mathcal{L}_{ij}^{ee}/d\tau$ stands for the (differential)
photon luminosity distribution constructed out of the photon
densities $f_{j/e_1}, f_{i/e_2}$ as follows:
\begin{eqnarray}
\frac{d\,\mathcal{L}_{ij}^{ee}}{d\tau} &=&
\int_\tau^1\,\frac{dx}{x}\,
\frac{1}{1+\delta_{ij}}\,\left[f_{i/e_1}(x)\,f_{j/e_2}(\tau/x) +
f_{j/e_1}(x)\,f_{i/e_2}(\tau/x)\right] \label{eq:photondensity}.
\end{eqnarray}
Functions $f_{i/e_1}$ and $f_{j/e_2}$ (one per beam of given
polarization) are taken, in our case, from the standard package
CompAZ~\cite{compaz}.

In order to proceed with the numerical analysis, let us first of all
examine the structure of the partonic cross-section. At fixed
$\hat{s}=M_h^2$, it can be expressed from Eq.\,(\ref{eq:generic2})
in a simply manner:
\begin{equation}\label{partonicsigma}
\sigma(\pdirect)=\frac{8\pi}{M_h^2}(1+\eta_1\,\eta_2)\,\mathcal{B}(h\to\gamma\gamma)\,.
\end{equation}
If averaged over polarizations, the cross-section is given by the
previous result but without the factor $1+\eta_1\,\eta_2$, because
one has to sum over $\eta_1,\eta_2=\pm 1$ and divide by $4$. For
polarized photon beams of equal polarization (i.e. $++$ or $--$),
the resulting cross-section is a factor of 2 larger. If,
alternatively, we consider the case of opposite polarizations
($+-$), the cross-section vanishes (as expected from angular
momentum conservation).

Let us emphasize that the partonic cross-section encodes already the
relevant information regarding the dynamical features of the 2HDM
under study. This information is obviously contained in the reduced
amplitude $\mathcal{M}$ of the effective $h\,\gamma\gamma$ vertex,
and thus also in the branching ratio of $h\to\gamma\gamma$. Notice
that, in the region where the triple interaction dominates, the
amplitude behaves roughly as
$ \mathcal{M} (\Pphoton\Pphoton \to \hzero) \sim {\alpha_{em}\,
C_{\PHiggs^+\PHiggs^-\hzero}}$.  Obviously, in this region, the
cross-section (\ref{partonicsigma}) is directly sensitive to the
$\PHiggs^+ \PHiggs^- \hzero$ trilinear Higgs self-interaction (cf.
e.g. the first two diagrams of the second row in Fig.
\ref{fig:loop}), given by
\begin{eqnarray}
C_{\PHiggs^+\,\PHiggs^-\hzero}
&=&\frac{i\,e}{2\,M_W\,\sin\theta_W}
\bigg[\sin(\beta-\alpha)\,
\left(M^2_{\hzero} - 2\,M_{\Hpm}^2\right)\nonumber\\
&&-\left.\frac{\cos(\beta+\alpha)}{\sin\,2\beta}\,\left(2\,M^2_{\hzero}
- 4\frac{1}{e^2}\,\lambda_5\,M_W^2\,\sin^2\theta_W \right) \right]
\label{eq:3h}.
\end{eqnarray}
It follows from this expression that such coupling can be enhanced
either at low or high values of $\tan\beta$, and also through the
Higgs boson mass splittings -- unlike the MSSM case. In addition, it
may be heightened through its explicit dependence on the $\lambda_5$
parameter\,\,\footnote{We note that for $\lambda_5 =
\lambda_6=2\sqrt{2}\,G_F\,M_{\Azero}^2=e^2\,M_{A^0}^2/(2\sin^2\theta_W\,M_W^2)$,
Eq.\,(\ref{eq:3h}) reduces to the corresponding result of Table 1 of
Ref.\,\cite{Ferrera:2007sp}, as it should. }. Needless to say, in
all cases these enhancements are strictly harnessed by the various
theoretical and phenomenological constraints discussed in the
previous section. By comparison, the corresponding (approximate)
behavior of the amplitude for the SM contribution, if we assume that
is dominated by the top quark Yukawa coupling, reads
 $\mathcal{M} (\Pphoton\Pphoton \to \HSM)
 \sim\,{N_c\,Q_t^2\alpha_{em}\,e\,m_t^4}/
({M^2_{H}\,M_W\,\sin\theta_W})$. This amplitude receives in general
sizeable contributions from the gauge boson loops that significantly
correct it. Similarly, in the domains of the 2HDM parameter space
where the 3H-coupling (\ref{eq:3h}) is not overwhelming over the
Yukawa and gauge boson couplings (cf. e.g. the first, third and
subsequent rows of diagrams in Fig. \ref{fig:loop}), our estimate
above does not even hold as a crude approximation. Therefore, in
general it is necessary to come to grips with the full expression
for the effective coupling $g_{\gamma\gamma h}$ (which is indeed the
main object under study, mainly for the cases $h=h^0, H^0$). Even
more useful is to define the exact ratio between the corresponding
2HDM and SM coupling strengths at one-loop:
\begin{equation}\label{ratior}
r\equiv\frac{g_{\gamma\gamma h}}{g_{\gamma\gamma
H}}=\frac{\big|\mathcal{M}\big|^{\rm
2HDM}}{\big|\mathcal{M}\big|^{\rm
SM}}=\left[\frac{\Gamma(h\to\gamma\gamma)}{\Gamma(H\to\gamma\gamma)}\right]^{1/2}\,.
\end{equation}
Obviously, the 2HDM amplitude will depend on whether the model is of
type-I or type-II. In the particular region where the trilinear
coupling dominates for $h^0$ production,
the expectation on that ratio is roughly of order $r\sim
C_{\PHiggs^+ \PHiggs^- \hzero}\,M_H^2\,M_W/m_t^4$. Therefore, in the
regime of large $|\lambda_5|$, for which $C_{\PHiggs^+ \PHiggs^-
\hzero}\sim M_{W}\, |\lambda_5|$, we can foresee a big enhancement
with respect to the SM. To be sure, in practice we will perform the
numerical analysis of the exact expression (\ref{ratior}) and
consider its behavior in general regions of parameter space.

For the sake of convenience, let us focus hereafter on the sets of
Higgs boson masses displayed in Table~\ref{tab:mass_sets} below.
\begin{table}[tbh]
\begin{center}
\begin{tabular}{|c||c|c|c|c|}
\hline
        2HDM & Set I & Set II & Set III & Set IV \\
\hline\hline
$M_{\hzero}$ & $115$ &  $150$ &  $200$  & $200$  \\
$M_{\Hzero}$ & $165$ &  $200$ &  $250$  & $250$  \\
$M_{\Azero}$ & $100$ &  $110$ &  $290$  & $340$  \\
$M_{\Hpm}$   & $105$ &  $105$ &  $300$  & $350$  \\
\hline
\end{tabular}
\end{center}
\caption{\footnotesize{\footnotesize{Higgs mass parameters, in GeV,
used throughout the calculation.}}} \label{tab:mass_sets}
\end{table}
The mass spectrum in Set I of Table~\ref{tab:mass_sets}, for
instance, allows to enhance the $\PHiggs^+\PHiggs^-\PHiggslightzero$
coupling. Its maximum value is roughly attained for $\sin\alpha=
-0.86$, $\tan\beta = 1.70$ and $\lambda_5=-25.0$. Incidentally,
notice that the aforementioned Set I of masses is only suitable for
type-I 2HDM, due to the relatively light value chosen for the
charged Higgs boson ($M_{\Hpm}=105$ GeV), which is below the
(indirect) limit of $295$ GeV afflicting the type-II
models\,\cite{Misiak:2006zs}. Let us recall that, experimentally,
the current $95\%$ C.L. direct mass limits for general Higgs bosons
searches are: $M_{\Hpm}\gtrsim 79.3$ GeV for the charged Higgs
boson, and $M_{h^0}\gtrsim 92.8$ GeV, $M_{A^0}\gtrsim 93.4$ GeV
($\tan\beta>0.4$) for the neutral ones (of course with
$M_{H^0}>M_{h^0}$)\,\cite{Amsler:2008zzb}.
\begin{figure}
\begin{center}
\includegraphics[scale=0.45, angle = -90]{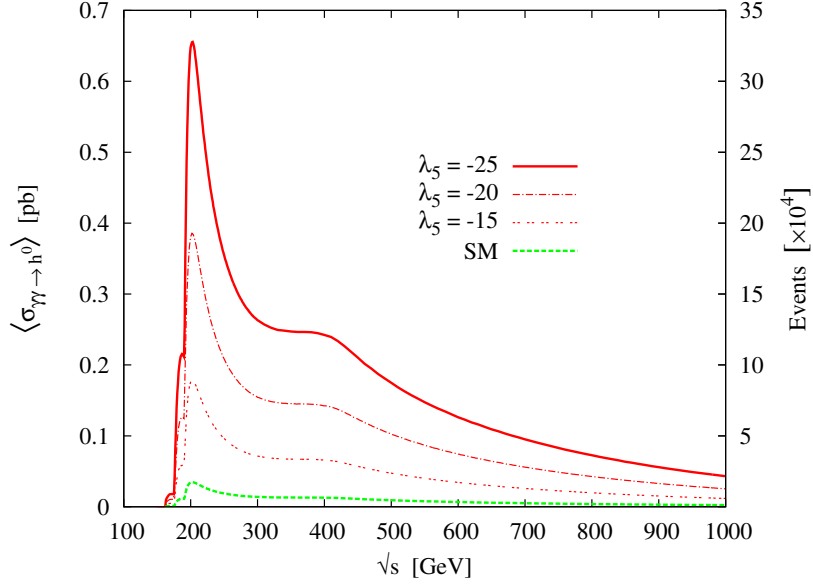}
\caption{\footnotesize{Cross-section $\langle\sigma_{\gamma\gamma\to
h}\rangle(s)$ given by Eq.\,(\ref{eq:sigmatotal}) and number of
Higgs boson events, as a function of the CM energy of the linac,
assuming an (integrated) luminosity $\mathcal{L}=500$ fb$^{-1}$. We
plot the corresponding values for the SM and the 2HDM using the Set
I of Higgs boson masses, $\sin\alpha=-0.86$, $\tan\beta=1.7$ and
three values for $\lambda_5$.}} \label{fig:eehsetI}
\end{center}
\end{figure}

In Fig.~\ref{fig:eehsetI}, we perform the numerical analysis of the
averaged total cross-section $\langle\sigma_{\gamma\gamma\to
h}\rangle(s)$, Eq.\,(\ref{eq:sigmatotal}), in which the partonic
contribution is folded with the effective luminosity function
(\ref{eq:photondensity}). We display $\langle\sigma_{\gamma\gamma\to
h}\rangle(s)$ as a function of the center-of-mass (CM) energy
$\sqrt{s}$ of the linac machine for the Set I of Higgs masses. In
this figure, we explore a region of parameter space where the
\pmzero coupling dominates for different negative values of
$\lambda_5$ (As we will see later on, large $\lambda_5>0$ values are
forbidden by vacuum stability.). Notice that, for sufficiently large
$|\lambda_5|>10$, the cross-sections can be considerably high
(spanning the range $0.01-0.2$ pb) at the fiducial startup value
$\sqrt{s}=500$\,GeV of the ILC, and entailing at this point more
than $10^3-10^4$ events for the given integrated luminosity. The
rates, however, decrease fast for smaller values of $|\lambda_5|$,
the reason being the destructive interference between the charged
Higgs, gauge boson and fermion loops at low values of $|\lambda_5|$.
For example, at the same energy and for $\lambda_5=(-2,-5,-8)$ {we
obtain $\langle\sigma_{\gamma\gamma\to h}\rangle= (2.40, 0.16,
5.50)$ fb} respectively.  Clearly, there is a delicate balance in
the low $|\lambda_5|$ region which amounts to a severe depletion of
the overall 2HDM cross-section. By comparison, the corresponding
result in the SM, namely for the same Higgs mass ($M_H=M_{h^0}$),
reads $\langle\sigma_{\gamma\gamma\to H}\rangle\simeq 11$ fb (hence
$\sim 5\times 10^3$ events at that energy and luminosity range),
which is quite sizeable. It follows that even small departures from
this value should be measurable, especially in a high precision
instrument as a $\gamma\gamma$ collider.
\begin{figure}[t]
\begin{center}
\includegraphics[scale=0.5]{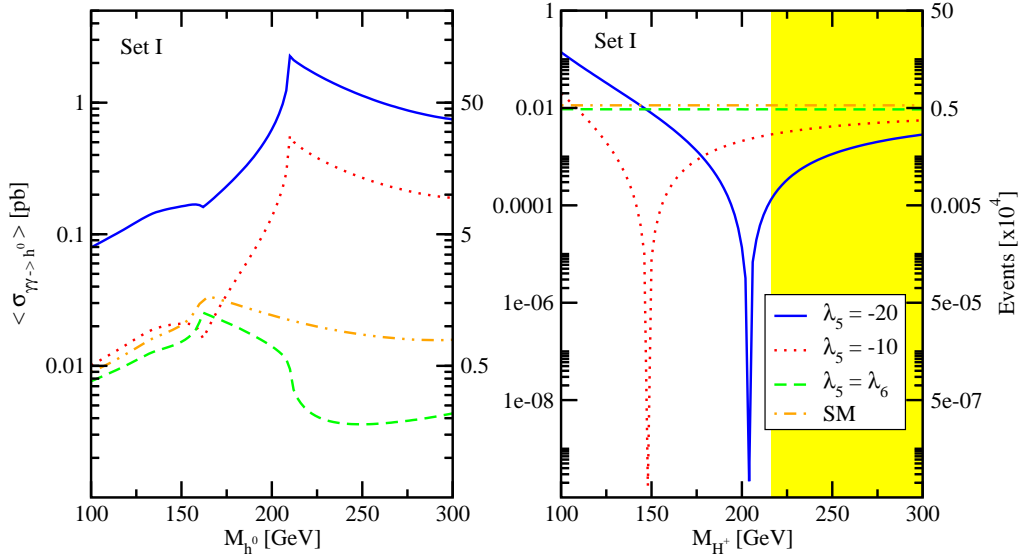}
\caption{\footnotesize{Cross-section $\langle\sigma_{\gamma\gamma\to
h^0}\rangle(s)$ as a function of the light CP-even Higgs mass (left
panel) and the charged Higgs mass (right panel) at fixed
$\sqrt{s}=500\,GeV$. Remaining mass parameters from Set I,
$\sin\alpha=-0.86$, $\tan\beta=1.70$ and three values for
$\lambda_5$: $-20$, $-10$ and $\lambda_5=\lambda_6$. The SM
cross-section $\langle\sigma_{\gamma\gamma\to h^0}\rangle$ is also
included (in the right panel, it almost coincides with the
$\lambda_5=\lambda_6$ case). The shaded area is excluded by the
constraints.}} \label{fig:overm}
\end{center}
\end{figure}

In Fig.\,\ref{fig:overm}, we test the dependence of the
cross-section $\langle\sigma_{\gamma\gamma\to h^0}\rangle$ on the
Higgs masses. Specifically, we plot its evolution in terms of
$M_{\hzero}$ (left panel) and $M_{\Hpm}$ (right panel). We take the
other parameters from Set I while keeping $\sin\alpha=-0.86$,
$\tan\beta=1.70$, and choose three different values for $\lambda_5$:
$-20$, $-10$ and
$\lambda_5=\lambda_6=2\sqrt{2}\,G_F\,M_{\Azero}^2\simeq 0.34$.
Included is also the SM cross-section
$\langle\sigma_{\gamma\gamma\to H}\rangle$. Worth noticing is the
fact that, while the 3H coupling is dominant, the cross-section does
not immediately drop with $M_{h^0}$; it actually increases for a
while up to a few hundred GeV. Moreover, for large values of
$|\lambda_5|$, the evolution of $\langle\sigma_{\gamma\gamma\to
h^0}\rangle$ presents a notorious ``spike-shaped'' enhancement near
$M_{\hzero}\sim 2\cdot M_{\Hpm}\sim 210$ GeV, which is brought about
by the threshold effect of two real charged Higgs bosons in the
loop. The corresponding effect for a pair of real vector bosons
$W^\pm$ at $M_{\hzero}\sim 2\cdot M_{W^\pm}\sim 160$ GeV is also
barely visible therein.

The sharp suppression dip standing out on the right panel of
Fig.~\ref{fig:overm} deserves also a few words. Recall that, in the
regime in which the influence of the trilinear contribution
(\pmzero) greatly ``waxes'', the charged-Higgs mediated correction
holds absolute sway over the loop-induced coupling $g_{\gamma\gamma
h^0}$. However, as soon as we raise the charged Higgs mass, the
positive influence of the trilinear Higgs boson interaction rapidly
wanes and it cancels more and more against the (negative) effects
from the gauge boson and fermion loops. This gives rise to the
aforementioned destructive interference. Eventually, a particular
value of $M_{\PHiggs^\pm}$ is reached where the two sorts of effects
virtually annihilate each other (right at the vertex of the dip in
the figure). Beyond this point, one rapidly reaches a regime where
only the gauge boson and Yukawa coupling (negative) effects remain.
Most of this region is actually excluded by the constraints. The
destructive interference pattern described here is only possible
when the set of Higgs boson masses is relatively light, as in the
case of Set I under consideration, otherwise the trilinear effects
could not be competitive.

The behavior of the effective coupling $g_{\gamma\gamma h^0}$ in the
2HDM can be better assessed in terms of the ratio $r$ defined in
(\ref{ratior}). Its dependence on the Higgs mass spectrum is sampled
in Table~\ref{tab:max} using the parameter setups indicated in
Table~\ref{tab:mass_sets}. Set II, for instance, contains a heavier
neutral CP-even Higgs sector which results in a sizeable value of
$r$ of $3.75$. It means that, in this case, the effective strength
of the $\gamma\gamma h^0$ vertex almost quadruples that of the SM
($\gamma\gamma H$). This is quite remarkable. On the other hand,
Sets III and IV are characterized by heavier charged Higgs bosons
and at the same time by a heavier CP-odd Higgs bosons (so as to
elude the $\delta\rho$ bounds). Unsurprisingly, the ratio $r$ falls
in this case to within values below $1$, i.e. close to the SM from
below. As it should be expected, the larger the charged Higgs mass
is, the less efficient is the enhancement capabilities associated to
the 3H self-interactions. Incidentally, let us notice that Sets III
and IV of Higgs-boson masses are intended to describe type-II 2HDM.
Does this mean that for type-II models (those closer to the MSSM
Higgs sector) there is no hope to hint at non-SM Higgs boson physics
with $\gamma\gamma$ collisions? Not necessarily so, as there is a
tail of subleading one-loop effects triggered by the non-SM Yukawa
couplings of the 2HDM in combination with the gauge bosons; in
particular, we have already detected it in Fig.\,\ref{fig:overm}b
for Set I (although within the excluded region). But, in general,
this tail is available and lies well within the allowed region for
heavier sets of Higss boson masses, such as Sets III and IV. More on
it below.

{A brief comparison with the existing calculations of single Higgs
boson production within the MSSM is in
order\,\cite{2phot_mssm,maria}. For example, in Ref.~\cite{maria}
the ratio between the decay widths of the MSSM Higgs boson $h^0$ and
the SM Higgs boson $H$ into $\gamma\gamma$ is computed at one-loop.
This ratio corresponds to the square of $r$ defined in
(\ref{ratior}). They find that, in the most favorable regimes (viz.
large mass-splitting and large mixing angle in the top squark
sector), it can lead to values of $r$ up to $\sqrt{2}\simeq 1.4$. It
follows that the most optimistic MSSM expectations on $r$ are
markedly below the maximum enhancement capabilities of the 2HDM ($r
\sim 4$). The reason for this is clear and it was already advanced
in the introduction -- namely, the 3H self-couplings within the MSSM
are restricted to be gauge-like and cannot source the potentially
large effects that we have identified within the general 2HDM.
Therefore, what we have called the ``tail of subleading effects'' in
the 2HDM case is actually one of the main sources of the MSSM
effects, the other being the supersymmetric Yukawa couplings of the
Higgs bosons with the squarks. }

\begin{table}
\begin{center}
\begin{tabular}{|c||c|c|c|c|} \hline
 & Set I & Set II & Set III & Set IV \\ \hline \hline
\bfseries $r$ &\bfseries 3.98 & \bfseries 3.75 & \bfseries 0.98 &
\bfseries 0.98 \\  \hline $\tan\beta$ & 1.7 & 1.7 &
1.0 & 1.0 \\ \hline $\sin\alpha$ & -0.86 & -0.86 & -0.82 & -0.82 \\
\hline $\lambda_5$ & -25 & -25 & 0 & 0 \\ \hline
\end{tabular}
\caption{\footnotesize{Maximum value of the ratio $r$,
Eq.\,(\ref{ratior}), in the case of $h^0$ and for the different mass
sets quoted in Table~\ref{tab:mass_sets}, together with the
configuration of $\tan\beta$, $\sin\alpha$, $\lambda_5$ for which
these optimal values are attained.}} \label{tab:max}
\end{center}
\end{table}

\begin{figure}
\begin{center}
\includegraphics[scale=0.44,angle=0]{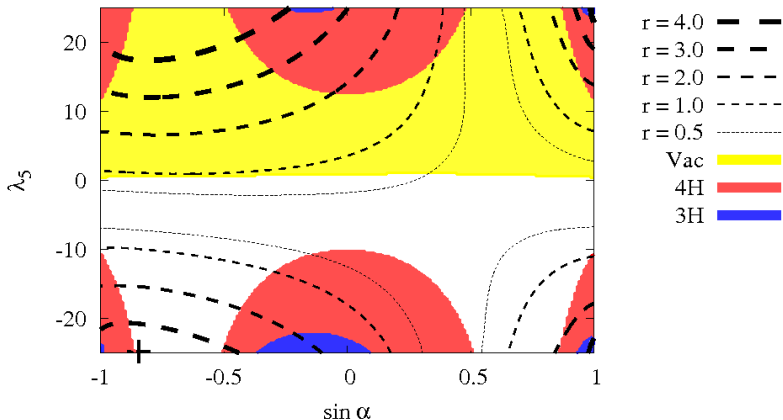}
\vspace{-1cm} \caption{\footnotesize{Contour lines for the ratio
$r={g_{\gamma\gamma h^0}}/{g_{\gamma\gamma H}}$ (\ref{ratior}) in
the $(\lambda_5, \sin\alpha)$ plane, for the Set I and $\tan\beta =
1.70$. The regions excluded by vacuum stability (upper half plane
$\lambda_5\gtrsim 0$) and the unitarity of the two kinds of Higgs
boson self-couplings, namely the quartic (large circular domains)
and the trilinear (small circular domains), are separately shown.
The cross symbol on the left (at the bottom) denotes the point with
maximum allowed $r$.}} \label{fig:scan1}
\end{center}
\end{figure}

Figure~\ref{fig:scan1} presents the corresponding contour lines for
the ratio (\ref{ratior}) in the $(\lambda_5, \sin\alpha)$ plane. The
cross symbol indicated on the left at the bottom denotes the point
in this plane where such ratio attains the maximum value permitted
by all the constraints. Moreover, we show the regions excluded by
vacuum stability and by unitarity of the trilinear and quartic Higgs
boson couplings. Notice that the sign $\lambda_5>0$ is mostly
forbidden by vacuum stability, which explains why we have presented
the numerical analysis of the previous figures only for
$\lambda_5<0$. In any case, we see that there is a sizable region
left where the effective 2HDM coupling $g_{\gamma\gamma h^0}$ is
significantly larger (in fact, a few times larger) than the SM
coupling $g_{\gamma\gamma H}$.

\begin{figure}[t]
\begin{center}
\includegraphics[scale=0.38, angle = -90]{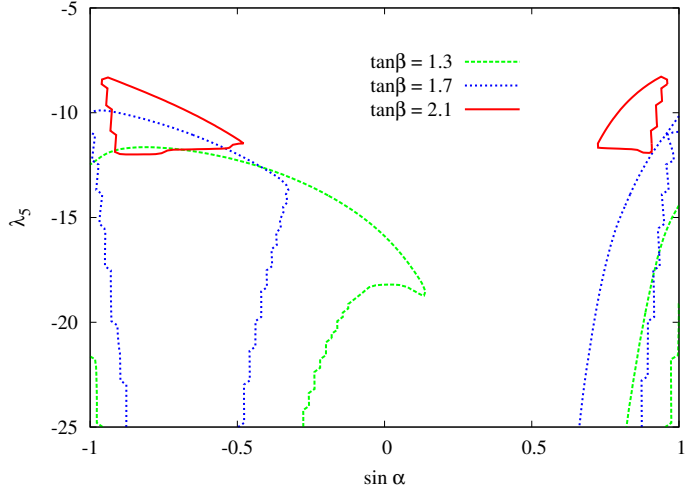}
\caption{\footnotesize{Regions in the $(\lambda_5, \sin\alpha)$
plane (allowed by all the constraints) where the ratio
(\ref{ratior}) is $r>1.1$, i.e. when $g_{\gamma\gamma h}$ is at
least $10\%$ bigger than $g_{\gamma\gamma H}$ for $\tan\beta=1.3,
1.7$ and $2.1$}} \vspace{0.0cm} \label{fig:silueta}
\end{center}
\end{figure}

Finally, the available domains across the $(\lambda_5, \sin\alpha)$
plane wherein one can obtain enhanced values of the ratio $r$ with
respect to the SM case is explored systematically in
Fig.~\ref{fig:silueta} for different values of $\tan\beta$ and at
the fiducial startup energy $\sqrt{s}=500$ GeV of the ILC.
Specifically, in this figure we compute those regions in which the
predicted value of the ratio $r$ exceeds the corresponding SM value
($r=1$) by $10\%$ at least while still being compliant with the full
set of constraints; equivalently, regions where the cross-section is
augmented by $20\%$ or more, thus inducing an excess of about one
thousand events above the SM prediction (within the given luminosity
segment). As we can see, the projected domains are sizeable.
Interestingly enough, even tiny $\pm 1\%$ departures of $r$ from $1$
would already be significant, as they would amount to $100$ events
up or down the SM expectation. If we would adopt this softer
criterion, the allowed domains in Fig.~\ref{fig:silueta} would be
substantially augmented.

Particularly stringent is the impact of the unitarity restrictions,
which translates into a progressive shrinking of the relevant region
as we raise the value of $\tan\beta$.
For completeness, we have also addressed the single production of a
heavy CP-even and a CP-odd Higgs boson. Their respective maximum
cross-sections (at $\sqrt{s}=500$ GeV) for the specific case of Set
I read as follows: $\langle\sigma_{\gamma\gamma\to H^0}\rangle =
0.31\,\picobarn$ and $\langle\sigma_{\gamma\gamma\to A^0}\rangle =
1.9 \,\femtobarn$. The former is of the order of
$\langle\sigma_{\gamma\gamma\to h^0}\rangle$ itself, whereas the
latter is significantly smaller, the depletion being caused by the
absence of trilinear couplings of $A^0$ with the charged Higgs
bosons. Furthermore, we wish to emphasize the existence of a region
of parameter space where $\langle\sigma_{\gamma\gamma\to
H^0}\rangle$ and $\langle\sigma_{\gamma\gamma\to h^0}\rangle$ can
both be simultaneously sizeable. That region (e.g. for Set I) is
just the allowed domain in the down-right corner in
Fig.\,\ref{fig:scan1}. There we find $\langle\sigma_{\gamma\gamma\to
H^0}\rangle\simeq 0.2$ pb and $\langle\sigma_{\gamma\gamma\to
h^0}\rangle\simeq 0.05$ pb, entailing some $10^4-10^5$ events within
the standard luminosity range.  This possibility can be very
relevant, as it could be responsible for a potentially distinctive
2HDM signature which is unmatched in the MSSM.

We have also tested the enhancement potential of the 2HDM in the
domains of parameter space where the 3H-coupling (\ref{eq:3h}) is
not relevant and where the bulk of the contribution concentrates on
the Higgs-fermion Yukawa couplings in combination with the gauge
bosons. In the case of generic type-II models, there is no
enhancement at large $\tan\beta$ (unlike the supersymmetric case)
inasmuch as $\tan\beta$ is severely constrained by unitarity. In
general, due to the destructive interference between the diagrams
dominated by the 3H-coupling and the rest (fermion and gauge boson
loops), the enhancement capabilities of the Yukawa sector become
overshadowed. As a consequence, in such region we meet the following
situation: 1) There are still non-negligible domains in the 2HDM
parameter space where the cross-section departs remarkably from its
SM counterpart. However, in most cases the departure entails a
significant (e.g. $10\%$) reduction  of the cross-section with
respect to the SM; 2) type-I and type-II models become essentially
indistinguishable in that domain. This is a reflect of the fact that
the Higgs-top quark coupling (which has the same form in either
type-I and type-II models) drives the leading contribution in the
Yukawa sector, whereas the gauge boson contribution is common in
both types of models. Therefore, spotting a tail of non-SM effects
in this region could not distinguish the type of 2HDM. Still, the
missing number of events could certainly hint at the existence of a
smoking gun triggered by physics beyond the SM.

Let us close this section by briefly mentioning that we have also
surveyed the impact of another, more restrictive, set of unitarity
constraints\,\cite{unitarity_main}. We have found that, in the most
optimistic scenario for \PHiggslightzero production in \pp
collisions, the \pmzero coupling lies roughly a factor $2-3$ below
the largest value it can take under the current set of constraints
\eqref{eq:unitary3h} and \eqref{eq:unitary4h}. Correspondingly, the
enhancement with respect to the SM would be a factor $5-10$ times
milder, thus dwarfing the relevance of the main effects in some
regions of the parameter space. We point out, though, that the
unitarity restrictions proposed in\,\cite{unitarity_main} are not
fully coincident with those considered in \cite{unitarity2}, and in
this sense there is still some controversy in the literature on this
issue.

\section{Single Higgs boson production through $\twophoton$ fusion}

The interest on virtual photon-fusion (``two-photon processes'')
certainly has a long and widespread history in Particle Physics. For
instance, a prospect for the measurement of the pion lifetime from
\begin{equation}\label{2ph}
\HepProcess{\APelectron\Pelectron \HepTo\Pphoton^*\Pphoton^* \HepTo
\APelectron\Pelectron +\pi^0}
\end{equation}
was first discussed by F. Low almost half a century ago\,
\cite{Low60}. Studies of these processes, as well as detailed
considerations on two-photon production of muon pairs and multi-pion
final states, were carried out subsequently in the
seventies\,\cite{Terazawa73}. Remarkably enough, long after the
first pioneering studies appeared, two-photon processes are still an
active and fruitful field of investigation, in particular for Higgs
boson production. In actual fact, single Higgs boson production is,
in a sense, the modern counterpart of Low's ``single-meson''
production from quasi-real two-photon collisions. Already in the
early eighties, this Higgs production channel was first studied in
the literature within the context of the old $e^+e^-$ (pre-LEP)
colliders by J.A. Grifols and R. Pascual\,\cite{GrifolsPascual80}.

To be sure, the traditional two-photon processes are the forerunner
of the future $\gamma\gamma$ colliders considered in the previous
section. These colliders will probably concentrate most of the
interest around linac physics in the future and may greatly
supersede the former in all practical searches for new physics. It
is instructive to see once more why, specially in regard to the
``Higgs issue'', a most sensitive matter these days. To this end, we
compute here the cross-section for the processes
\begin{eqnarray}
\HepProcess{\APelectron\Pelectron \HepTo \Pphoton^*\Pphoton^* \HepTo
\APelectron\Pelectron + h} \ \ \ \ \ \ (h =
\PHiggslightzero,\PHiggsheavyzero,\PHiggspszero) \label{eq:p2}\,,
\end{eqnarray}
and compare with the results for real $\gamma\gamma$ collisions
studied in the previous section. In practice, we concentrate on the
lightest CP-even state. Noteworthy is the fact that the
cross-section for the virtual $\gamma^{*}\gamma^{*}$-fusion
processes, in contradistinction to the real $\gamma\gamma$
collisions, grows with the CM energy up to very high values of
$\sqrt{s}$. Generically, the behavior of $\APelectron\Pelectron \to
\gamma^*\,\gamma^*\to Y+ \APelectron\Pelectron$ in the asymptotic
energy regime goes as
$\sim(\alpha^4/M^2)\,\ln^2(s/m_e^2)\,\ln^n(s/M^2)$, where $M$ is the
threshold mass of the produced final state $Y$, and the number
$n\geq 1$ depends on the high energy behavior of
$\sigma(\gamma\gamma\to Y)$. The logarithmic growth simply tracks
the dynamical feature by which, for these processes, the virtual
photons $\Pphoton^*$ can be quasi-real and hence have their momenta
well-below the CM energy of the process, which may satisfy $s \gg
M^2_V$ -- the rest of the energy being carried away by the
concomitant lepton final states. Recently, the pairwise production
of Higgs-bosons via weak gauge-boson fusion mechanism,
$\APelectron\Pelectron \to V^*V^*\to h\, h \,+ X\ (V=W^{\pm},\, Z;\
\ h = \PHiggslightzero,\PHiggsheavyzero,\PHiggspszero,\Hpm)$
was analyzed in Ref.~\cite{Hodgkinson:2009uj} at the leading order
$\mathcal{O}(\alpha^4_{ew})$. The process turned out to be
instrumental for probing the 3H self-interactions. A complementary
strategy along the same lines is offered by the
$\gamma\gamma$-fusion process (\ref{eq:p2}), in which the
loop-induced $\Pphoton\Pphoton h$ vertex at order
$\mathcal{O}(\alpha^4_{em}\,\alpha_{ew})$ can be dominated by the 3H
coupling.
\begin{figure}
\begin{center}
\includegraphics[scale=0.39, angle = -90]{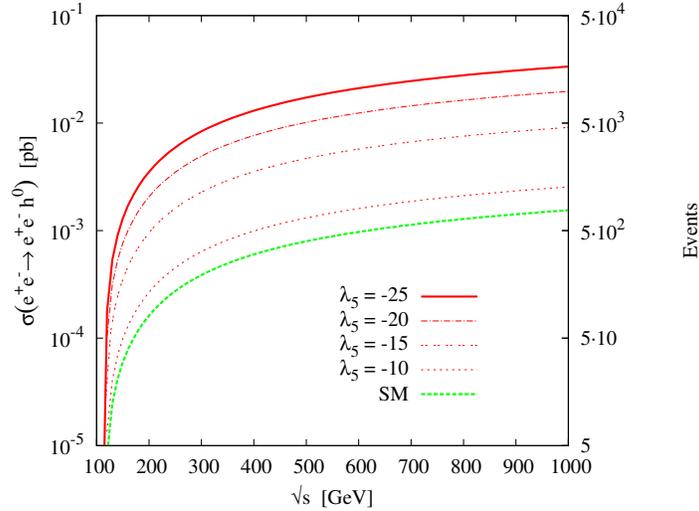}
\caption{\footnotesize{Evolution of the two-photon cross-section
$\sigma(\pminifusionh)$ as a function of the CM energy and the
corresponding number of events for a total luminosity
$\mathcal{L}=500$ fb$^{-1}$. We include the 2HDM and SM curves using
the Set I mass parameters, $\sin\alpha=-0.86$, $\tan\beta=1.7$ and
four values of $\lambda_5$.}} \label{fig:epa}
\end{center}
\end{figure}

The technical complications associated to two-photon processes of
the kind (\ref{2ph})-(\ref{eq:p2}) were tackled long ago in the
literature (cf. \cite{Terazawa73} for a classical review). The
physics is also well understood, it boils down to the well-known
\emph{equivalent-photon or Weizs\"acker-Williams approximation}, by
which the virtual photon emitted from the scattered electron appears
near the mass shell. This feature allows to essentially trade the
process of electro-production for a photo-production one with an
appropriate photon spectrum. The so-called \emph{photon} content of
a given electron can be explicitly factorized and the production
cross-section can be approximated in the following way:
\begin{eqnarray}
\sigma(\APelectron\Pelectron \to \APelectron \Pelectron X)
&=& \left[\frac{\alpha_{em}}{2\pi}\,\log\left(\frac{s}{4\,m_e^2}\right) \right]^2\,
\int_{\tau_0}^1\,f(\tau)\,\sigma_{\Pphoton\Pphoton \to X}\,(\tau\,s)\,d\tau
\label{eq:depa1},
\end{eqnarray}
where $\tau_0\equiv M_X^2/s$ and $f(\tau) =
({1}/{\tau})\,\left[(2+\tau)^2\,\log({1}/{\tau})
 - 2\,(1-\tau)(3+\tau)\right]$.

Figure \ref{fig:epa} presents the logarithmic evolution of the
production cross-section $\sigma(\pminifusionh)$ as a function of
the CM energy. We have used the mass Set I and the optimal
parameters quoted in Table~\ref{tab:max}. Once the production
threshold has been surpassed, the cross-section suddenly increases
up to a value of about $10^{-2}$ pb and becomes persistently
sustained, with only a mild (logarithmic) evolution. Over this
approximate plateau, it amounts to more than $5000$ events for an
integrated luminosity $\mathcal{L}=500$ fb$^{-1}$, therefore
resulting in an enhancement of one order of magnitude with respect
to the SM. Even more remarkable is the fact that, despite the fairly
large value of the two-photon cross-section, it is still a factor of
$10$ (at least) smaller than that of the corresponding
$\gamma\gamma$ real scattering processes, for the same values of the
2HDM parameters (cf. Fig\,\,\ref{fig:eehsetI}). This should explain
vividly and convincingly the outstanding superiority of the future
$\gamma\gamma$ colliders versus the ancient two-photon processes.

\section{Conclusions}

We have devoted this work to analyze the production of a single
neutral Higgs boson, $h =
\PHiggslightzero,\,\PHiggsheavyzero,\,\PHiggspszero$, within the
two-Higgs-doublet model (2HDM) through the following complementary
mechanisms: i) direct scattering of real photons in a
$\Pphoton\Pphoton$ collider; and ii) fusion of a virtual photon pair
in a conventional two-photon process, $\pfusion$. Both mechanisms
are direct handles on the effective $\vertex$ interaction,
$g_{\gamma\gamma h}$. This coupling is a pure quantum effect
generated by a plethora of radiative corrections involving charged
Higgs bosons, quarks and gauge bosons. Among the interactions in the
loops, we have the 3H self-interactions, most remarkably \pmzero in
the case of $h^0$ production -- which proves to be utterly dominant
in certain regions of the 2HDM parameter space. We have
systematically swept this space and identified those configurations
for which the departure from the SM prediction is most remarkable,
and we have done this in full compliance with the rigorous
constraints dictated by perturbativity, unitarity and vacuum
stability bounds, as well as by the EW precision data. The result is
that, in the most favorable scenarios, the physical cross-section
(i.e. the one convoluted with the backscattered luminosity function)
can typically reach the level of $\langle\sigma_{\gamma\gamma\to
h}\rangle\sim 0.1-1\,\picobarn$. In other words,
$\langle\sigma_{\gamma\gamma\to h}\rangle$ may rocket to values
$10-100$ bigger than the expected SM yield
$\langle\sigma_{\gamma\gamma\to H}\rangle$ (for similar values of
the Higgs mass), which by itself should already be perfectly
measurable: $\langle\sigma_{\gamma\gamma\to H}\rangle\sim 0.01$ pb
$=10$ fb. Such notorious enhancement can be traced back to the
behavior of the \pmzero coupling, despite it is highly restrained by
the overall constraints. By sticking to moderate $\tan\beta\gtrsim
1$, it is possible to licitly increase the rates by choosing
relatively large (negative) values of the parameter $\lambda_5$.
Moreover, in order to optimize this mechanism, the charged Higgs
boson should be relatively light (say, below $300\,\GeV$) so that
the associated quantum corrections are not severely hampered by the
decoupling effects. It means that the $g_{\gamma\gamma
h}$-enhancements that we have encountered apply only for type-I
2HDM, because for this kind of models the mass of the charged Higgs
boson is not constrained by $\bsg$. A similar conclusion ensues for
the case of double Higgs production in $\gamma\gamma$ collisions,
but with lower cross-sections\,\cite{2phot_2hdm}.

In the above conditions, the expected number of single Higgs boson
events emerging from direct $\gamma\gamma\to h$ scattering, within
the typical energy range of the ILC ($500 - 1000\,\GeV$), is of the
order of $10^5$ per $500$ $\invfb$ of integrated luminosity.
Compared to the production rate of single Higgs bosons through the
traditional virtual photon-pair fusion, $\pfusion$ (which performs
at the level of $0.01\,\picobarn$, at most, for $\sqrt{s}\geq 500$
\GeV), the real $\gamma\gamma$-collision mechanism is at least one
order of magnitude more efficient.

On the experimental side, the prospects for Higgs boson detection in
a $\gamma\gamma$-collider are deemed to be excellent. To start with,
let us stress that the single Higgs-boson final state is to be
produced essentially at rest. Therefore, for $M_h<2M_V\lesssim
180\,GeV$, the corresponding signatures should mostly be in the form
of back-to-back, highly energetic, quark jets ($b\bar{b}$,
$c\bar{c}$). For $M_h>2M_V$, instead, signatures with two or four
charged leptons in the final state (from $W^{\pm}\to
\ell^{\pm}+\text{missing energy}$ and, specially, from $Z\to
\ell^+\ell^-$) should be really pristine. Furthermore, we have seen
that, in some cases, the two channels $\gamma\gamma\to h^0$ and
$\gamma\gamma\to H^0$ are simultaneously accessible and with similar
rates. Needless to say, this could result in a double distinctive
signature of new physics.

With enough statistics on these events, and upon analyzing the
invariant mass distribution of the resulting jets of quarks and
leptons, the measurement of some 2HDM Higgs boson(s) mass(es) should
be attainable with fairly good accuracy, together with a precise
determination of the effective $g_{\gamma\gamma h}$ couplings
(typically for $h^0$ and/or  $H^0$). If their strengths would happen
to be vastly dominated by the triple Higgs boson self-interactions,
the signature of non-standard Higgs boson physics would be
crystal-clear, leading  us to suspect it to be rooted in some
generic type-I 2HDM. However, should we meet the juncture
$g_{\gamma\gamma h}\lesssim g_{\gamma\gamma H}$ or $g_{\gamma\gamma
h}\gtrsim g_{\gamma\gamma H}$, the underlying quantum effects would
be largely insensitive to the type of model and a more detailed
comparative study with the MSSM would be mandatory\,\cite{BDS2}.
Even then, tiny deviations could hint at new physics. In this
regard, it is important to emphasize that, given the high precision
nature of a $\gamma\gamma$ collider, gathering a small $5-10\%$
effect (positive or negative) should be sufficient to point at a
smoking gun.

\vspace{0.25cm}
 \noindent
\textbf{Acknowledgments}\,\, JS is grateful to F. Mescia for useful
discussions on the bounds on the charged Higgs boson mass from low
energy physics. NB thanks an ESR position of the EU project RTN
MRTN-CT-2006-035505 Heptools; DLV acknowledges the MEC FPU grant
AP2006-00357. DLV and JS have been supported in part by MEC and
FEDER under project FPA2007-66665 and by DURSI Generalitat de
Catalunya under project 2005SGR00564. This work was partially
supported by the Spanish Consolider-Ingenio 2010 program CPAN
CSD2007-00042.

\newcommand{\JHEP}[3]{ {JHEP} {#1} (#2)  {#3}}
\newcommand{\NPB}[3]{{\sl Nucl. Phys. } {\bf B#1} (#2)  {#3}}
\newcommand{\NPPS}[3]{{\sl Nucl. Phys. Proc. Supp. } {\bf #1} (#2)  {#3}}
\newcommand{\PRD}[3]{{\sl Phys. Rev. } {\bf D#1} (#2)   {#3}}
\newcommand{\PLB}[3]{{\sl Phys. Lett. } {\bf B#1} (#2)  {#3}}
\newcommand{\EPJ}[3]{{\sl Eur. Phys. J } {\bf C#1} (#2)  {#3}}
\newcommand{\PR}[3]{{\sl Phys. Rept. } {\bf #1} (#2)  {#3}}
\newcommand{\RMP}[3]{{\sl Rev. Mod. Phys. } {\bf #1} (#2)  {#3}}
\newcommand{\IJMP}[3]{{\sl Int. J. of Mod. Phys. } {\bf #1} (#2)  {#3}}
\newcommand{\PRL}[3]{{\sl Phys. Rev. Lett. } {\bf #1} (#2) {#3}}
\newcommand{\ZFP}[3]{{\sl Zeitsch. f. Physik } {\bf C#1} (#2)  {#3}}
\newcommand{\MPLA}[3]{{\sl Mod. Phys. Lett. } {\bf A#1} (#2) {#3}}
\newcommand{\JPG}[3]{{\sl J. Phys.} {\bf G#1} (#2)  {#3}}

\end{document}